\documentclass[vecphys]{svmult}

\usepackage{graphicx}        
\usepackage[bottom]{footmisc}

\begin{document}

\title*{Radio and Infrared observations of Transition Objects}
\author{L. Cerrigone\inst{1,2},
J. L. Hora\inst{1}, G. Umana\inst{2} \and C. Trigilio\inst{2}}
\institute{Harvard-Smithsonian Center for Astrophysics \\
60 Garden St \\
Cambridge, MA, USA \\
\texttt{lcerrigone@cfa.harvard.edu}
\and INAF - Catania Astrophysical Observatory \\
via S. Sofia 78 \\
Catania, Italy \\}
%
%
\maketitle

\begin{abstract}
Observing objects in transition from pre- to young Planetary Nebula (PN), when the central star radiation starts to excite the envelope, can help us to understand the evolution of the circumstellar ejecta and their shaping mechanism/s. In our project we have selected a sample of hot post-AGB stars as Transition Phase candidates. 
Radio observations 
have led to detect free-free radiation from an ionized shell in about half of our targets, 
providing us with two sub-samples of ionized and non ionized Transition Objects.
We are now using IRAC and IRS on the Spitzer Space Telescope to determine if extended  emission is present (IRAC) and to study our targets' chemistry (IRS). In particular, by comparing spectra from the two sub-samples, the IRS observations will enable us to check how the presence of an ionization front effects the circumstellar envelope. The IRAC measurements, combined with previous ones in the literature, will give us information on the extent and physical conditions of the dust components.
\end{abstract}

\section{Introduction}
Planetary Nebulae (PN) evolve from intermediate mass stars. During their evolution they go through the Asymptotic Giant Branch (AGB) and then into the PN phase, ending their days as white dwarfs. The formation and early evolution of PN are a poorly understood phase of stellar evolution. In particular, it is not clear yet how the spherically symmetric AGB circumstellar shells transform themselves into the non spherical symmetries observed in the envelopes of evolved PN. Many observational programs have been devoted to recognize new planetary and proto-planetary nebulae among unidentified IRAS sources with far infrared colors similar to those of known PN \cite{pottasch88, parthapottasch} with the final aim of understanding PN formation through the discovery and analysis of new Transition Objects (TO). In these sources the physical processes associated with PN formation, such as dynamical shaping, are still occurring and, as the number of known transition objects is extremely small, the identification and study of new samples will be very important for further testing current models of stellar evolution. 

To detect new TO we have selected a sample of stars classified as post-AGB on the basis of their IRAS colors and optical spectra. We have selected our stars from lists of post-AGB candidates available in the literature \cite{parthasarathy, suarez}. The main selection criteria were  IRAS colors typical of PN and spectral classification as B stars. The B spectral type implies the central star can be hot enough to have started to ionize the circumstellar envelope. This maximizes the chance that the objects  are in the transition from post-AGB to PN. These criteria led to the selection of 36 targets.


\section{Observations}
{\bf Radio detections}\\
To define the evolutionary status of the selected sources  checking for the presence of ionized gas in their envelopes, we have observed all of the selected targets with the Very Large Array 
(VLA). Observations were performed in two runs, one in 2001 and another one in 2005 \cite[]{umana}. The observations in 2005 were performed in dynamical mode in June and July. The array was in CnB configuration in 2001 and in C and CnB in 2005.
All the observations were performed at 8.4 GHz, as the VLA receivers in this band are more sensitive than the others and the compact configurations were necessary to detect possibly low surface brightness objects. This setup enabled us to detect radio emission in 16 targets out of the 36 selected, for a 44\% detection rate. 
The detection of radio emission is  proof of the ionization of the envelope and therefore confirms the evolutionary status as TO/very young PN.
~\\
~\\
{\bf \noindent Infrared Observations}\\
As a follow-up to our radio detections we have observed our sample with IRAC and IRS onboard the Spitzer Space Telescope.
In general, comparisons of IR images of PN, which trace the molecular and warm dust emission, to optical  line images, which trace the ionized gas, have shown the presence of similar structures \cite{latter}. This leads to the conclusion that molecular and ionized gas spatially coexist in these sources, as well as dust grains, despite the different physical conditions these components are presumed to survive in. Our IRAC observations have two goals: detecting the presence of extended emission from the remnant AGB envelope and determining the Spectral Energy Distribution to check for the presence of different dust components. The IRS observations aim at revealing the chemistry of the envelopes.
Observations are  planned for 29 of our objects. All IRAC observations have been performed, while all but three targets have been observed with IRS.

\section{Results and Discussion}
{\bf IRAC and IRS observations}\\
IRAC has proved to be a very sensitive instrument, revealing previously unseen structures in the outer shells of expanded PN \cite{hora}. Since the nebulae in our sample are typically very compact and IR-bright, which is normal for their evolutionary status, their core emission dominates the inner few arcsec near the central source. This unfortunately usually prevents us from detecting any weak extended emission near the core, but, by PSF subtraction,  does not prevent us from measuring our targets' fluxes. Given the size of the PSF, an upper limit of $\sim5''$ to the sizes of the nebulae can be estimated.

\begin{figure}[htbp]
\hspace{-0.2cm}
\begin{minipage}{6cm}
\centering
\includegraphics[width=5.5cm]{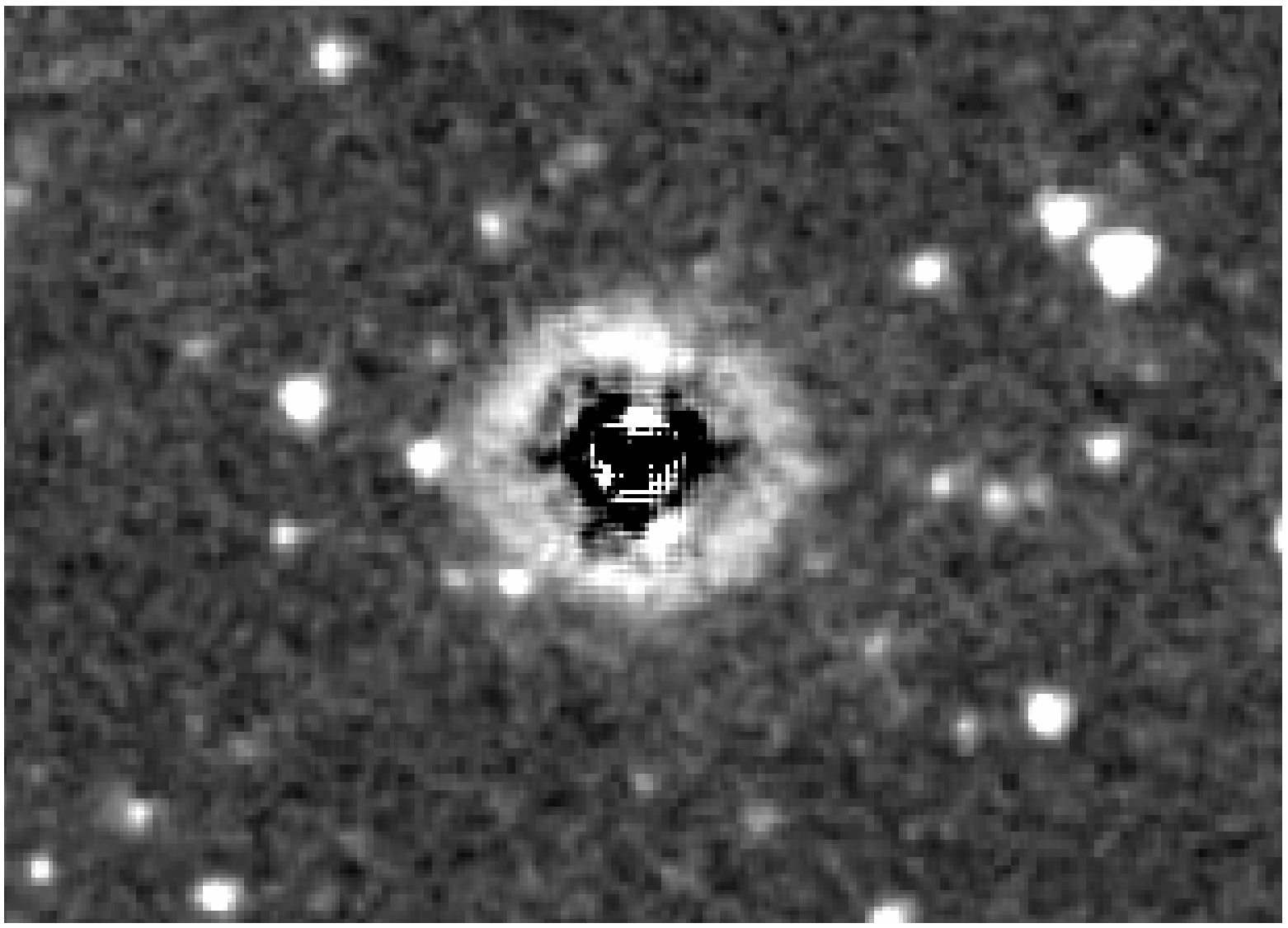}
\end{minipage}
\begin{minipage}{6cm}
\centering
\includegraphics[width=5.5cm]{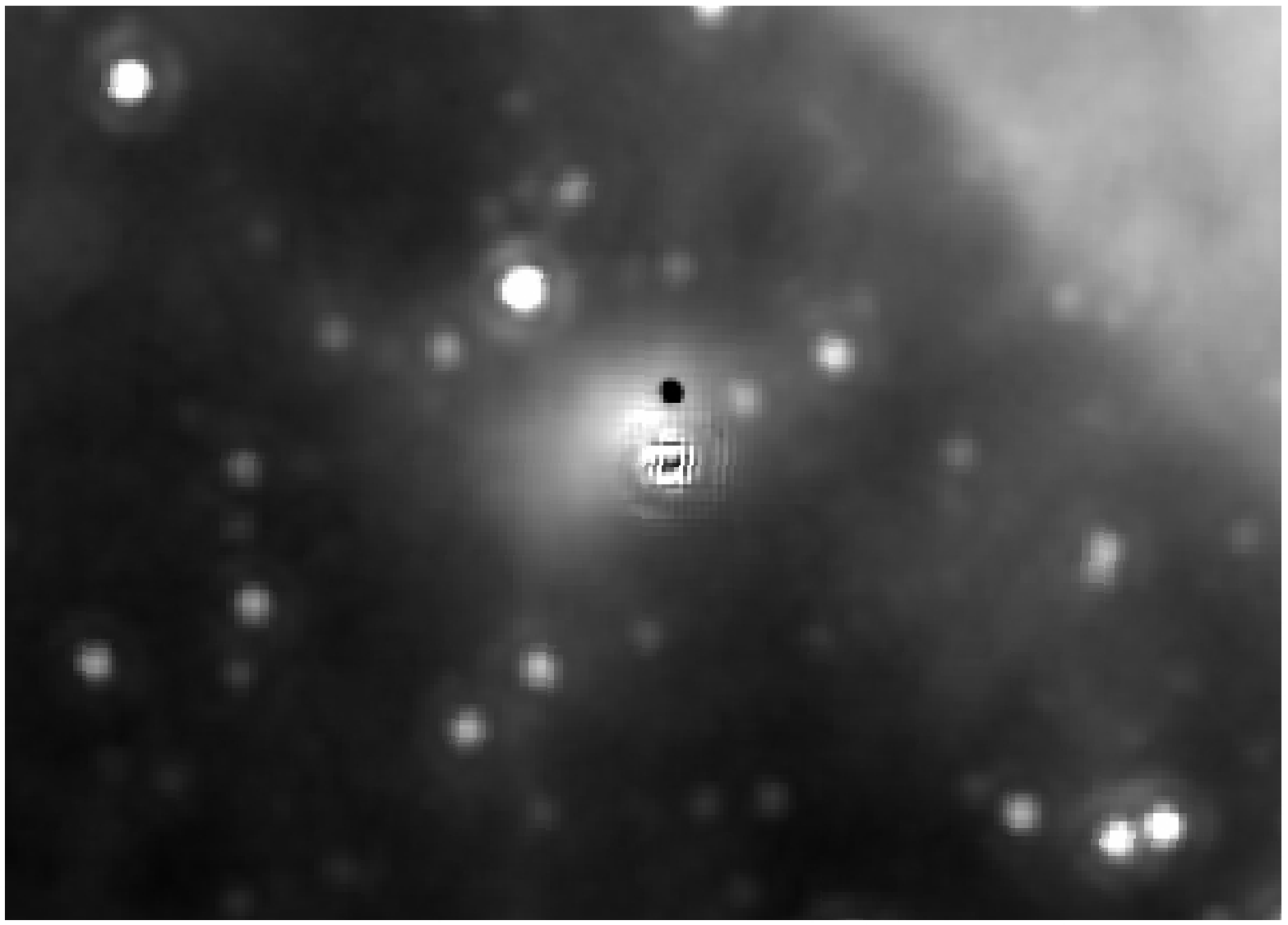}
\end{minipage}
\caption{{\it Left}: IRAC PSF subtracted image of IRAS 19590-1249 at 3.6 $\mu$m, where a ring with $\sim20''$ diameter surrounding the central star. {\it Right}: PSF subtracted image at 8.0 $\mu$m of IRAS 18070-2346 with an asymmetrical distribution of material around the central object.}
\label{fig:irac}
\end{figure}

In two targets we have clearly detected extended emission: IRAS 18070-2346 and IRAS 19590-1249 (Figure~\ref{fig:irac}). 
In I19590 the ring surrounding the central star is seen in channel 1 and only very weakly in channel 2. This is probably explainable in terms of dust reflecting radiation coming from the central star. Since no emission from this ring is seen in channel 3 or 4, it is likely made up of cold dust, whose emission peaks  beyond the mid-IR.

As previously mentioned we have also planned a spectral follow up with IRS, that has been performed at low resolution (R$\sim$90), covering the 5--38 $\mu$m range, with S/N$\sim$50. 
\begin{figure}[htbp]
\centering
\includegraphics[width=8cm]{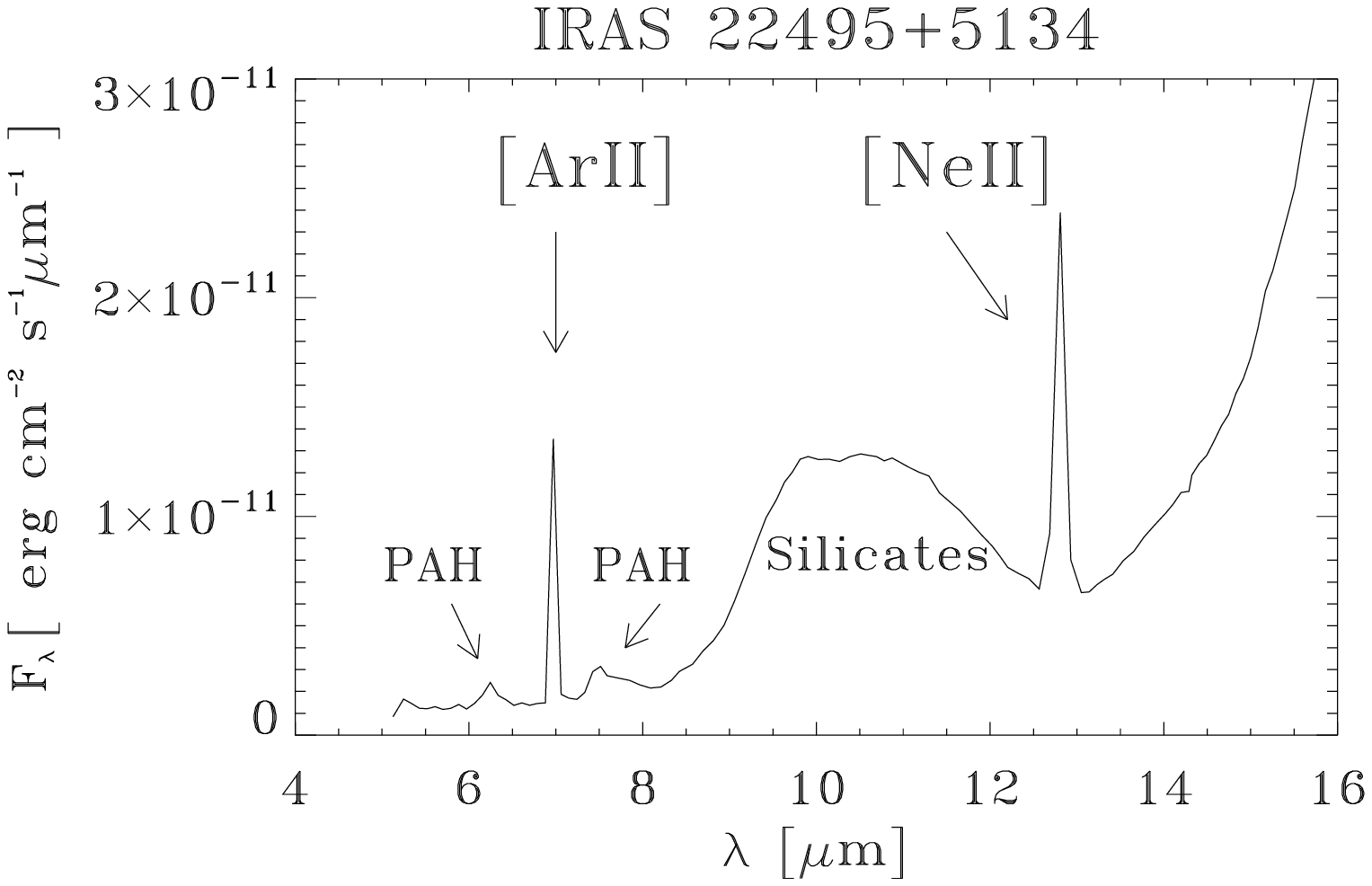}
\includegraphics[width=8cm]{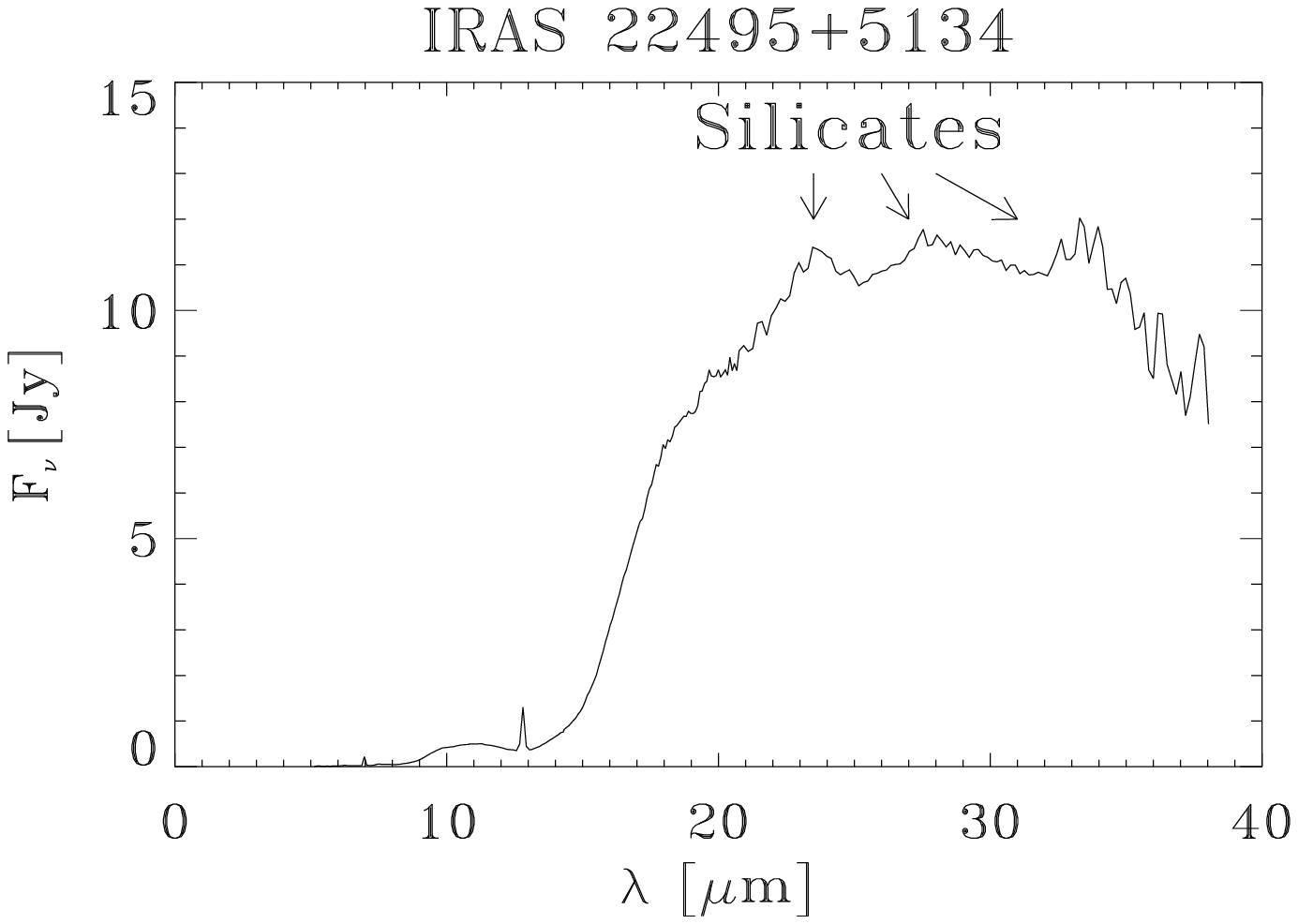}
\caption{IRS spectrum of IRAS 22495+5134, showing the presence of crystalline silicate features beyond 20 $\mu$m ({\it bottom}) and both PAH and silicate features at short wavelengths ({\it top}).}
\label{fig:irs22495}
\end{figure}
Figure~\ref{fig:irs22495} shows our spectrum for IRAS 22495+5134. The crystalline silicate features around 23.7, 27.6 and 33.6 $\mu$m and a blending of several features around 10 $\mu$m are clearly visible and can be attributed to forsterite \cite{molster}. At short wavelengths we can also identify some features due to Polycyclic Aromatic Hydrocarbons (PAH) at 6.2 and 7.7 $\mu$m and two features from ionized gas: [Ar II] at 6.98 $\mu$m and [Ne II] at 12.8 $\mu$m. Figure~\ref{fig:irs20462} shows the spectrum of another target in our sample, IRAS 20462+3416. Like in the previous one, both silicate and PAH features are detected. In this star we also identify the 11.3 $\mu$m PAH feature, while no [Ar II] is detected and [Ne II] is weaker than in the previous object. The ionized gas features detected in these objects are indicators of a higher excitation level in I22495 than in I20462. Both targets have been detected in our radio observations.
\begin{figure}[htbp]
\centering
\includegraphics[width=8cm]{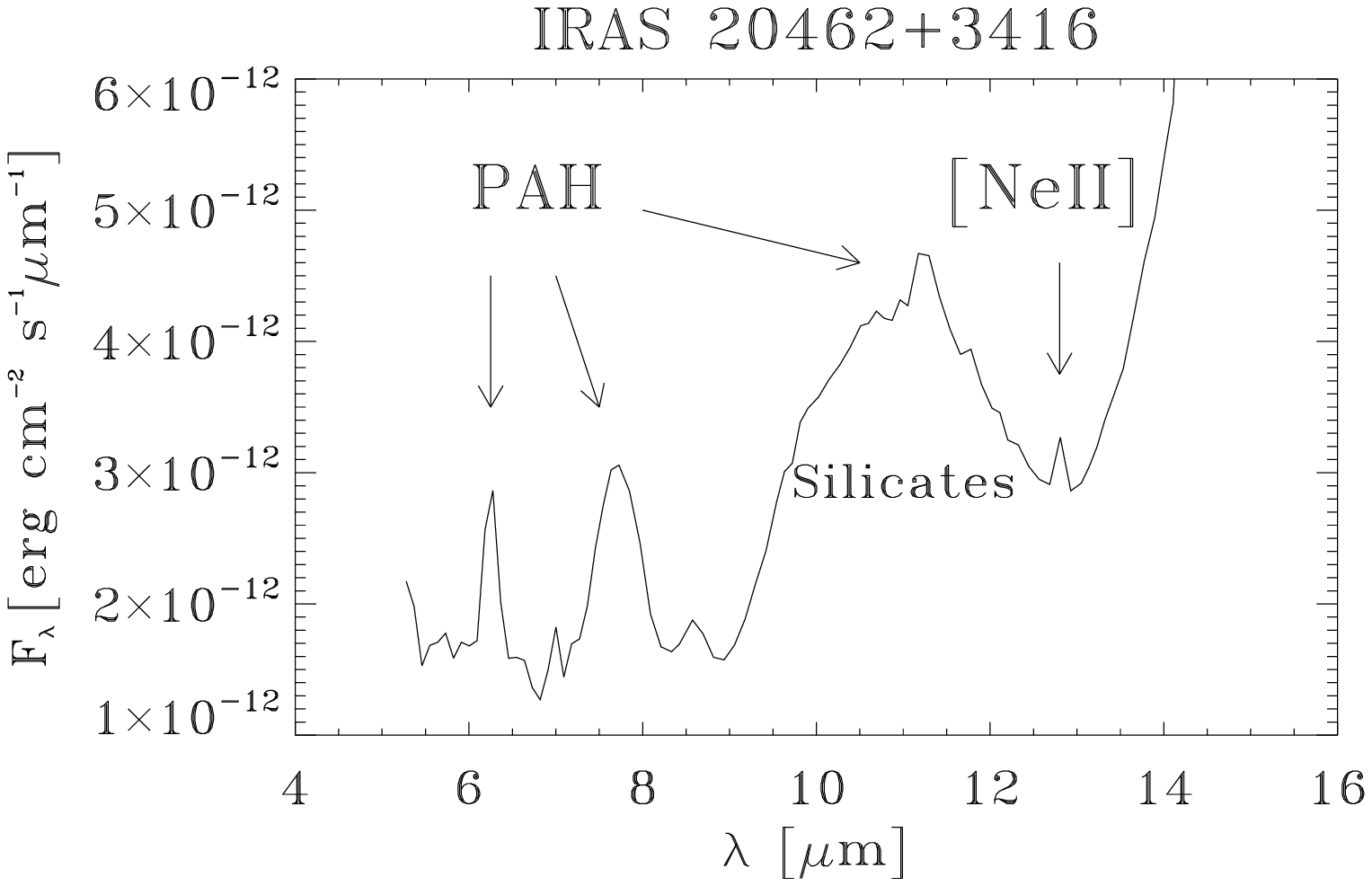}
\includegraphics[width=8cm]{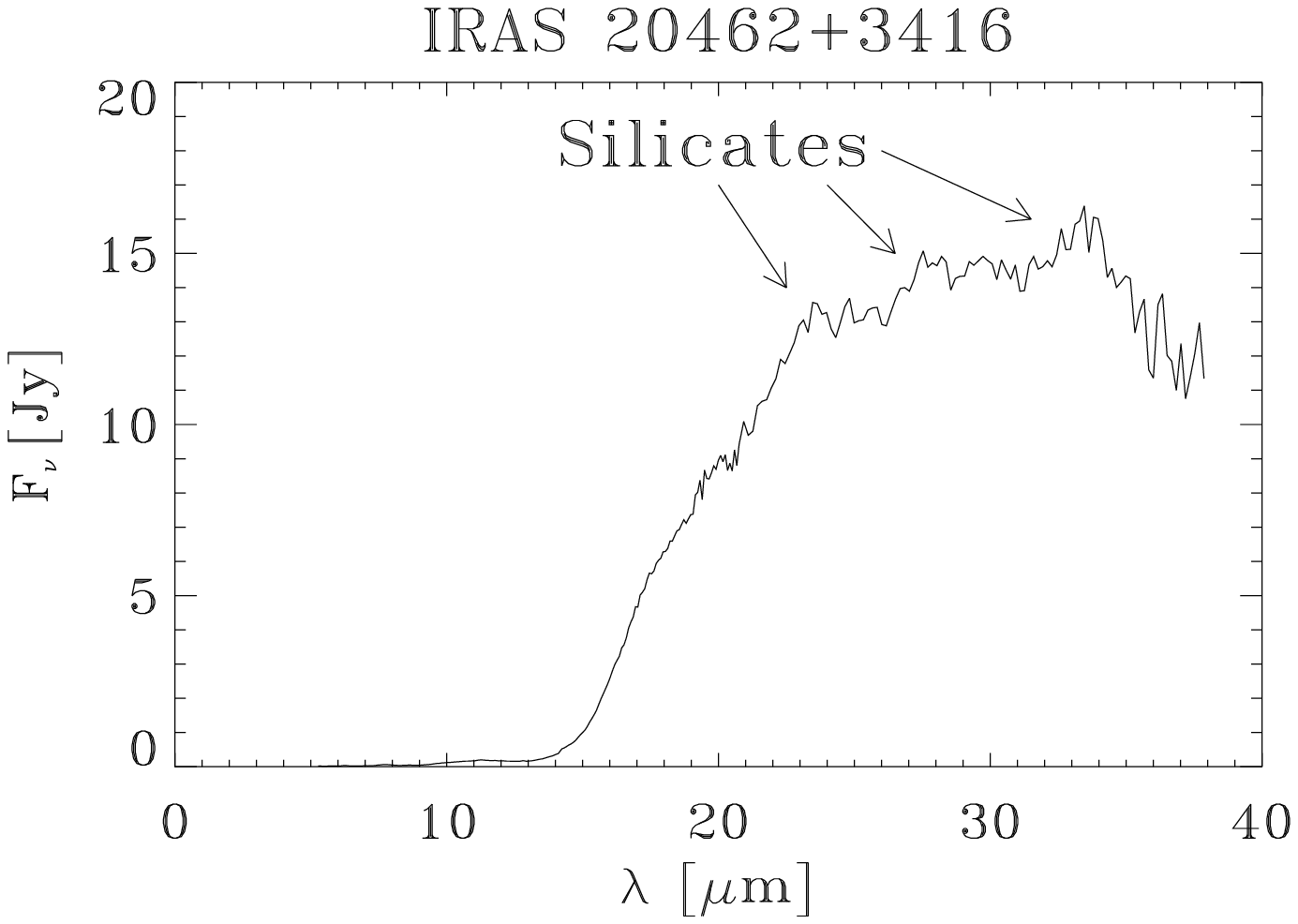}
\caption{Like in Figure~\ref{fig:irs22495} but for IRAS 20462+3416.}
\label{fig:irs20462}
\end{figure}

PN are usually classified as C-rich or O-rich. This classification is linked to the evolution of the central star, because it depends on the possibility that the star goes through a third dredge-up. It is this event that alters the chemistry in the envelope turning it from O-rich into C-rich during the AGB phase \cite{salaris}. Because of the stability of the CO molecule, if C is less abundant than O, all C is trapped in CO and then the envelope shows features of molecules containing oxygen (besides CO) and viceversa if O is less abundant than C.

It is known however that a minority of stars shows {\it mixed chemistry}, with molecules containing carbon and molecules containing oxygen. The origin of this mixed chemistry is not yet well understood. A noteworthy result from our TO study is that about 50\% of the stars in our sample show both PAH and silicate features. Among these mixed chemistry objects about 50\% are radio detected nebulae and 50\% are not. One question is whether the presence of PAH features is indicative of a mixed chemistry or these molecules can be somehow produced also in O-rich environments \cite{volk}. An explanation for the mixed chemistry involves the presence of disks, where an O-rich chemistry would be preserved from the third dredge up, while elsewhere in the envelope features from C-rich molecules would arise. \\
~\\
{\bf \noindent Spectral Energy Distribution}\\
We have searched the IRAS and 2MASS archives and retrieved data for our targets. 2MASS data were dereddened \cite{schlegel} and were combined with our IRAC observations to build up the Spectral Energy Distribution (SED) of the sample stars. We then tried to reproduce the SED by modelling the sources with the DUSTY code \cite{dusty}, which solves the radiation transfer equation in dust rich environments. It can use either spherical or slab geometry, and several O-rich and C-rich chemical components can be introduced. The code needs as an input the properties of the central source of radiation, the envelope density, grain size distribution, temperature of the inner dust layer, shell thickness relative to the inner radius and the optical depth at an arbitrary wavelength. 

In our attempts to reproduce the observed data points, the following inputs were used: a blackbody central source, density varying as $r^{-2}$, grain size distribution varying  as $a^{-3.5}$ with $a_{min}=0.005$, $a_{max}=0.26$ and optical depth at 60 $\mu$m calculated from the IRAS data \cite{gathier}. 

\begin{figure}[htbp]
\centering
\includegraphics[width=8cm]{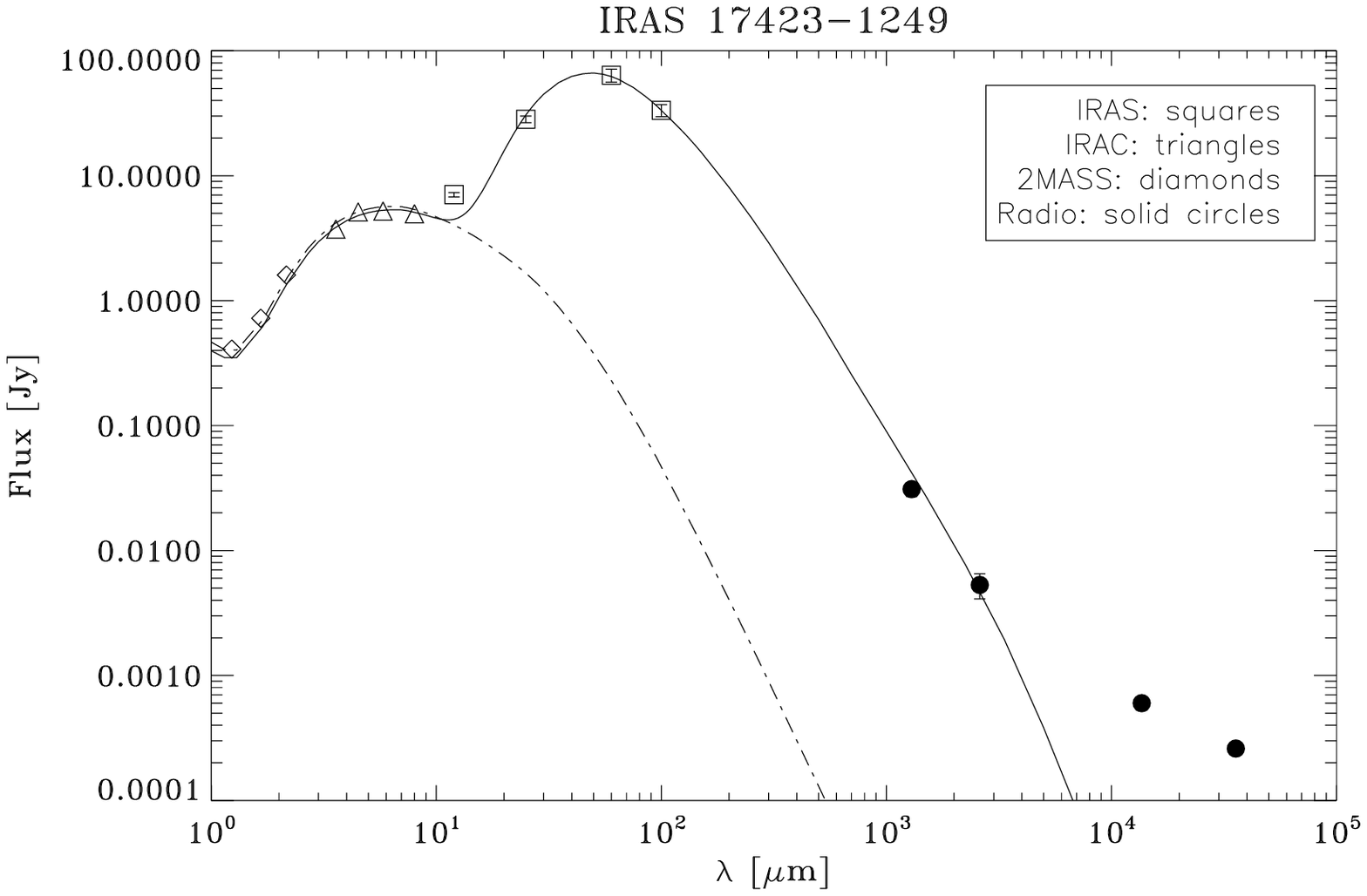}
\includegraphics[width=8cm]{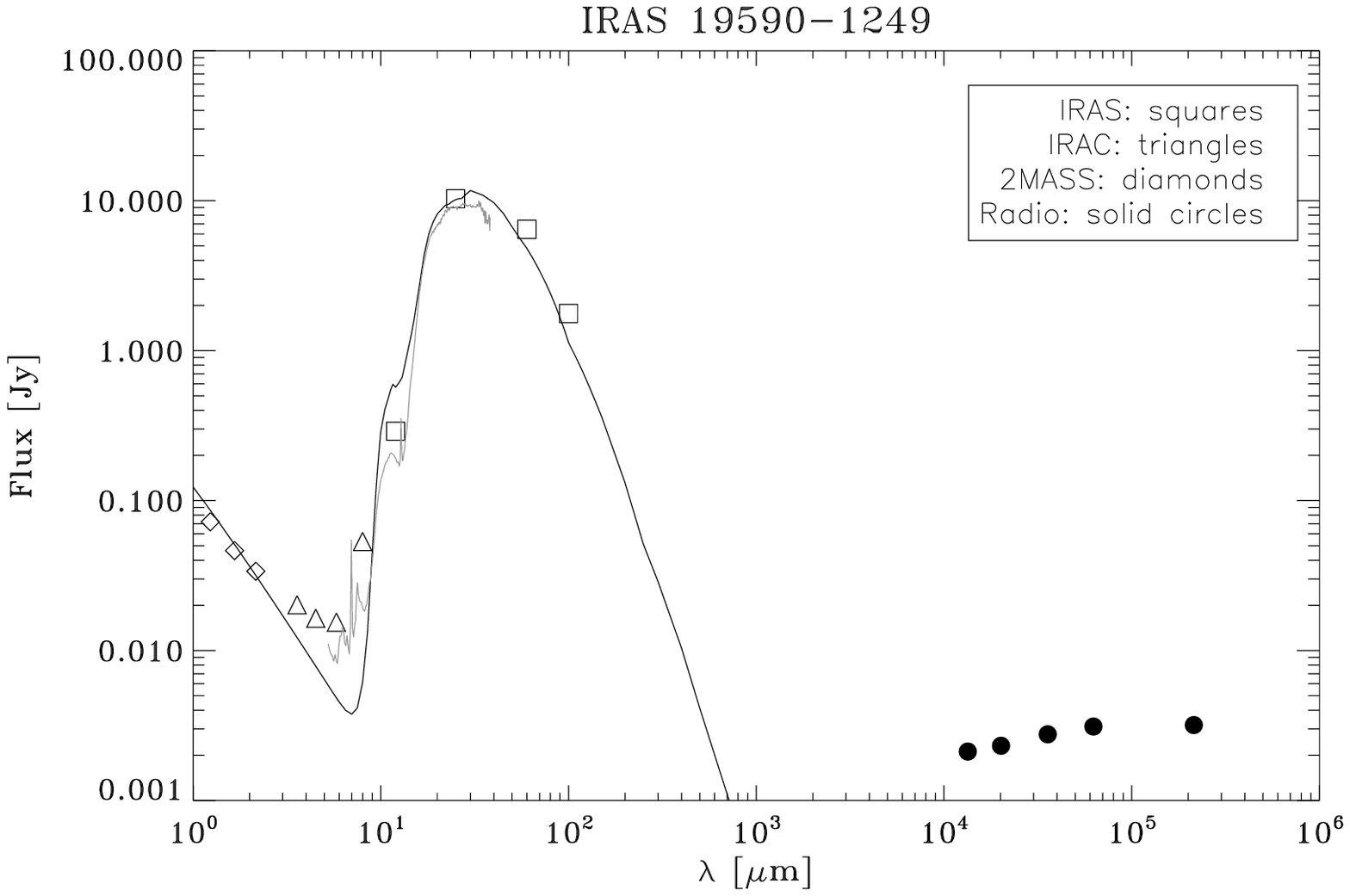}
\caption{DUSTY modelling of the SED for two of our targets: IRAS 17423-1755 ({\it top}) and IRAS 19590-1249 ({\it bottom}). For I17423 we have overplotted the single component at 1150 K (dot-dashed curve). For I19590 we have overplotted the IRS spectrum.}
\label{fig:dusty}
\end{figure}

The DUSTY modelling allows us to identify the dust components in the envelopes and estimate their temperatures. Figure~\ref{fig:dusty} shows the results of this procedure in two examples: IRAS 17423-1755 and IRAS 19590-1249.
For IRAS 17423-1755 we are not able to reproduce the data with only one component. It is necessary to have  dust at 1150 K and at 110 K, with the hot component having grain sizes from 0.005 to 4.7 $\mu$m and the cold one from 0.04 to 100 $\mu$m. Large grains are necessary to account for the mm emission \cite{huggins}. We have scaled our model and calculated the optical depth assuming a size of 1.8$''\times1.1''$ (which is the size of the CO shell), distance 5.8 kpc, and luminosity 12600 L$_\odot$ \cite{huggins}. 
We have used a chemistry with amorphous carbon only, since the ISO spectrum is almost featureless with the exception of a broad absorption feature at 3.1 $\mu$m, due to C$_2$H$_2$ and/or HCN \cite{gauba}. 

For IRAS 19590-1249 one component only, at 130 K, is sufficient to reproduce the observations with silicate dust properties. If we assume the distance of 4 kpc and luminosity 6300 L$_\odot$, we basically confirm the previous modelling performed \cite{bogdanov}. We have assumed a maximum size for the dust shell of $5''$, because a larger structure would have been observed in our IRAC images. Our model does not take into account the outer ($\sim$ $10''$ radius) ring, whose effects seem negligible. This is in agreement with the assumption that it is made up of cold dust, then its contribution to the observed flux, probably larger in the sub-mm or mm range, is negligible in the observations performed so far. Table~\ref{tab:dusty} summarizes the input used for both targets.
\begin{table}[htbp]
\centering
\begin{tabular}{lccc}
\hline\noalign{\smallskip}
    &       \multicolumn{2}{c}{IRAS 17423}  & IRAS 19590         \\
\noalign{\smallskip}\hline\noalign{\smallskip}
T$_\star$ (K) &   \multicolumn{2}{c}{20000}    & 23750       \\
Chemistry  &            \multicolumn{2}{c}{amorphous C}  & Silicates \\
Density    & \multicolumn{2}{c}{$r^{-2}$} & $r^{-2}$  \\
T$_d$  (K)    &   1150     & 110    & 124 \\
$a_{min}$ ($\mu$m) & 0.005   &  0.04 &    0.005\\
$a_{max}$ ($\mu$m) & 4.7     &  100     &  0.26  \\
$\tau_{60}$ &   0.002  &    0.03     & $2.63\times10^{-4}$       \\
Y          &   723                           & 4.55               & 4.1\\ 
Distance  (kpc) & \multicolumn{2}{c}{5.8} & 4 \\
Luminosity (L$_\odot$) & \multicolumn{2}{c}{12600} & 6300 \\
\noalign{\smallskip}\hline
\end{tabular}
\caption{Parameters used in modelling two of the targets.}
\label{tab:dusty}
\end{table}

\section{Conclusions}
We are in the process of characterizing a sample of post-AGB  objects, whose infrared colors and optical spectra are typical of Transition Objects. The detection of radio continuum emission in almost half of our sample is a confirmation of such an evolutionary status.
IRS observations have so far indicated that a mixed chemistry is more common than in other PN samples and subsequent observations are necessary to investigate the nature of this emission. IRAC observations, along with 2MASS and IRAS archive data, allow us to search for the presence of more than one emitting component in the dust envelope of the selected sources and model them to estimate such parameters as temperature and  grain sizes. 

{\acknowledgement L. Cerrigone acknowledges funding from the Smithsonian Astrophysical Observatory through the SAO Predoctoral Program. This work is based in part on observations made with the Spitzer Space Telescope, which is operated by the Jet Propulsion Laboratory, California Institute of Technology under a contract with NASA. Support for this work was provided by NASA through an award issued by JPL/Caltech. The Very Large Array is a facility of the National Radio Astronomy Observatory, which is operated by Associated Universities, Inc., under cooperative agreement with the National Science Foundation.}

%
%

%
%

%

%
%
%
%



\end{document}